\documentstyle[epsfig]{mn} 
\title[The mass of the black hole of Cygnus A]
{Spectroscopy of the near-nuclear regions of Cygnus A:
estimating the mass of the supermassive black hole}

\author[Tadhunter et al.]
       {C. Tadhunter$^{1}$, A. Marconi$^{2}$, D. Axon$^{3}$, K. Wills$^{1}$, T.G. Robinson$^{1}$, N. Jackson$^{4}$
	\\
$^{1}$Department of Physics and Astronomy, University of Sheffield,  Sheffield, S3 7RH, UK\\ 
$^{2}$Osservatorio Astrofisico di Arcetri, Largo E.Fermi 5, I-50125, Firenze, Italy \\
$^{3}$Department of Physics, Rochester Institute of Technology,
85 Lomb Memorial Drive, Rochester, New York 14623-5603, USA \\
$^{4}$Jodrell Bank Observatory, University of Manchester, Macclesfield, Cheshire, SK11 9DL. \\}

\date{}
\def\ltsim{\ifmmode\stackrel{<}{_{\sim}}\else$\stackrel{<}{_{\sim}}$\fi}
\def\gtsim{\ifmmode\stackrel{>}{_{\sim}}\else$\stackrel{>}{_{\sim}}$\fi}
\begin{document}
\maketitle
\begin{abstract}{\large We use a combination of high spatial
resolution optical and near-IR spectroscopic data 
to make a detailed study of the kinematics
of the NLR gas in the near-nuclear regions of  the  powerful, 
FRII radio galaxy Cygnus A ($z=0.0560$), with
the overall goal of placing limits on the mass of any supermassive black hole
in the core. Our K-band infrared observations (0.75 arcsec seeing) 
-- taken with NIRSPEC on the
Keck II telescope -- show a smooth rotation pattern across
the nucleus in the Pa$\alpha$ and
H$_2$ emission lines along a slit position (PA180) close to perpendicular to the
radio axis, however, there is no evidence for such rotation along the radio
axis (PA105). Higher spatial resolution observations of the 
[OIII]$\lambda$5007 emission
line -- taken with STIS on the Hubble Space Telescope (HST) -- confirm
the general rotation pattern of the gas in the direction perpendicular to the 
radio axis, and provide evidence for steep velocity gradients within a radius
of 0.1 arcsec of the core --- corresponding to the high surface brightness
structure visible in high resolution narrow band images. The 
[OIII] line remains broad throughout the core region ($FWHM \sim 300$ --
$900$ km s$^{-1}$), but the spatial distribution of [OIII]
provides no evidence for an unresolved inner narrow line region (INLR).
Assuming that the ionized gas is circularly rotating in a
thin disk and that the large line widths are due to activity-induced 
turbulence, 
the circular velocities measured from both the Keck and HST data
lead to an estimate of the mass of the supermassive black hole
of $2.5\pm0.7\times 10^9$M$_{\odot}$. For the host galaxy
properties of Cygnus A, this mass is consistent with the global correlations
between black hole mass and host galaxy properties
deduced for non-active galaxies.
Therefore, despite the extreme power of its radio source and
the quasar-like luminosity of  its AGN, the black hole in
Cygnus A is not unusually massive considering the
luminosity of its host galaxy. Indeed, the estimated mass of the black hole in Cygnus A is similar to that
inferred for the supermassive black hole in the FRI
radio galaxy M87, despite the fact
that the AGN and radio jets of Cygnus A are 2 -- 3 orders of magnitude more powerful.
Overall, these results are consistent with the idea that the properties
and powers of the AGN in radio galaxies are determined as
much by the mass 
accretion rates, as by the absolute masses of their supermassive black holes.    

As well as providing evidence for a supermassive black hole
in the core of Cygnus A, our data also demonstrate that nuclear activity 
has an important effect on
the kinematics of the circum-nuclear gas on a sub-kpc scale. 
Most notably, the velocity offsets measured
in the two outer HST/STIS slit positions are consistent with
the presence
of an activity-induced outflow in the NW cone. 
}

\end{abstract}
\newpage
\section{Introduction}

With the high spatial resolution afforded by the 
Hubble Space Telescope (HST) and
high resolution ground-based instruments there has been considerable
progress over the last decade in determining the central mass 
distributions
of galaxies. In consequence, 
there is now compelling evidence for massive dark objects
(MDOs) in the cores of many apparently ``normal'' galaxies. Moreover, the 
masses of the MDOs are found
to correlate with both the luminosities and the velocity
dispersions of the bulges of the host galaxies (e.g. Kormendy
\& Richstone 1995, Gebhardt et al. 2000, Ferrarese \& Merritt 2000).

Given that supermassive 
black holes are generally considered to be the driving force
for much of the extreme activity observed in active
galactic nuclei (AGN) and the accompanying radio-emitting jets, 
there is clearly an interest in extending
studies of the MDOs to active galaxies and quasars. Such studies
are important because the properties
of the putative central supermassive 
black holes and the associated accretion flows
are likely to be crucial in determining the properties
of the AGN, including:
the total power output; the presence or absence of an accretion disk;
the shapes of the SEDs; and the ability to form powerful radio
jets.

However, kinematic studies of the mass distributions in the central regions of
active galaxies face several potential difficulties. These include:
\begin{itemize}
\item {\bf Resolution.} Most luminous AGN are relatively distant, and it can 
be difficult to resolve the central regions in sufficient detail 
to determine the accurate mass distributions.
\item {\bf Radial motions.} Intense nuclear activity may be accompanied by
outflows in the warm emission line gas driven by jets, AGN-driven winds or
circum-nuclear starbursts. Such outflows may distort the velocity patterns
in the near-nuclear regions.
\item {\bf Extinction.} Dust extinction associated with the central obscuring
tori and/or kpc-scale dust lanes, may ``hide'' some of the circum-nuclear
structures orbiting around the black hole, and lead to a failure to observe the
full range of velocities in the circum-nuclear disks.
\item {\bf Contamination by the AGN}. Contamination
by the psf of the bright AGN can
hamper attempts to measure gas and stellar kinematics in the near-nuclear
regions (depending on orientation). 
\end{itemize}

Partly as a consequence of these difficulties, few active galaxies have direct
measurements of their central mass distributions, based on studies of the
kinematics of the narrow emission line gas. This problem is most acute
for radio-loud active galaxies: only a  few genuine radio
galaxies --- including M84, M87 and  Cen A --- 
have accurate measurements of their 
central mass distributions, and all of these belong to the low luminosity,
FRI, sub-class (e.g. Bower et al. 1998, Marconi et al. 1997, 2001). 

Based on indirect
black hole mass estimates, there have been suggestions 
that the radio properties of active galaxies may 
be linked to the 
mass and/or relative mass accretion rates 
of the central supermassive black holes (e.g. Franceschini, Vercellone
\& Fabian 1998;  McMclure et al. 2000; 
Ghisellini \& Celotti 2001).
Therefore, it is important to attempt to determine accurate
dynamical masses for the MDOs in
the central regions of more powerful, FRII, radio galaxies.

In this paper we present high spatial resolution observations of the
emission line gas in the central
regions of the nearby FRII radio galaxy Cygnus A ($z=0.0560$), 
taken with the aim
of determining the central mass distribution. 
The observations, which comprise near-IR observations taken with the 
Keck II telescope, as well as optical observations taken with the Hubble Space
Telescope (HST), illustrate both the potential and the pitfalls
of observations of this type. They allow us to estimate
the mass of the black hole in Cygnus A. They also provide important information
about the structure and kinematics
of the narrow line region (NLR) in this key object. 


We assume the cosmological parameters $H_0 = 75$ km s$^{-1}$ Mpc$^{-1}$
and $q_0 = 0.0$ throughout this paper. For these parameters 
1.00 arcsecond corresponds to 1.00 kpc at the redshift of Cygnus A.

\section{Previous observations of Cygnus A}

Cygnus A is the most powerful radio source in the local universe ($z < 0.5$)
and also one of the best observed across the electromagnetic spectrum. A
full review of the properties of Cygnus A is presented in Carilli \& 
Barthel (1996). To provide a context to the results presented below, in
this section we describe relevant observations at X-ray, optical and near-IR
wavelengths.

Much of the work over the last decade has involved using
Cygnus A to test the orientation-based unified schemes for powerful radio sources,
which suggest that all powerful radio galaxies contain hidden quasar
nuclei (Barthel 1989).

The most direct evidence for a hidden quasar  in Cygnus A
is provided by X-ray observations which show evidence for
a nuclear  power-law component
at hard X-ray energies (Ueno et al. 1994, Young et al. 2002), 
and optical imaging- 
and spectro-polarimetry observations which show a biconical reflection 
nebula (Tadhunter et al. 1990, Ogle et al. 1997) and scattered broad line emission lines in polarized light (Ogle et al. 1997). Although
a point-like nuclear source is detected at near-IR wavelengths
in this object (Djorgovski et al. 1991, Tadhunter et al. 1999), the high
polarization of this source suggests that it represents scattered light
from a near-nuclear reflection nebula, rather that the quasar nucleus
observed directly in transmission (Tadhunter et al. 2000). Based on
the absorption corrected X-ray luminosity of the core, 
and assuming that Cygnus A has
a spectral energy distribution similar to the radio-loud quasars
in Elvis et al. 1994, the AGN in Cygnus A has a bolometric luminosity
in the range $5\times10^{45} < L_{BOL} < 2\times10^{46}$ erg s$^{-1}$. This
places it at the lower end of the range of quasar bolometric luminosities. 

The presence of a luminous, hidden illuminating source is further supported
by optical narrow-band imaging observations which reveal
kpc-scale ionization cones both to the south east and north west of the radio core
(Jackson et al. 1998), and
long-slit spectroscopy observations which show a ``U''-shaped
ionization pattern in the direction perpendicular to
the radio axis on the NW side of the nucleus
(Tadhunter, Metz \& Robinson 1994). 
There is no evidence for an unresolved emission line
source in the nucleus at optical wavelengths, although a 
high-surface-brightness ``V''-shaped emission line structure is 
detected close to the position of 
the radio core (Jackson et al. 1998).

The kpc-scale cones and reflection nebulosity are also detected at near-IR
wavelengths (Tadhunter et al. 1999), where they appear edge-brightened close
to the nucleus. The edge-brightening
of the cones strongly suggests that they have been hollowed out by 
circum-nuclear outflows.

A major advantage of Cygnus A for black hole studies is that, provided the
rotation axis of the inner disk is parallel to the radio axis and to the
axis of the inner torus, the inclination ($i$) of this axis to the
line of sight is well-constrained. From the jet/counter-jet flux
ratio and apparent expansion speed of the VLBI jet, the
results of Sorathia et al. (1996), Krichbaum et al. (1998)
and Bach et al. (2002) are consistent with 
$40 < i < 89$ degrees for our adopted cosmology, depending
on the degree of  free-free absorption of the counter-jet by the
torus close to the nucleus (see discussion in Krichbaum
et al. 1998). If we assume
that the starburst ring detected in colour composite HST images
of the nuclear regions of Cygnus A is intrinsically circular
(Fosbury et al. 1999), the  
observed degree of flattening of the ring leads to $i = 57^{+8}_{-10}$.
Perhaps the strongest constraint, however, is provided by the 
that fact that we do not detect the quasar nucleus
directly at optical wavelengths and therefore 
cannot be looking directly into cones. Using the opening
half angle of the cones measured from the HST emission line
images ($55 < \theta_{1/2} < 60$ degrees) this
constraint leads to $i > 50$ degrees for the gas in the inner
(obscuring) regions
of Cygnus A. Overall, observations at both optical and radio wavelengths
are consistent with the idea that the radio jet is perpendicular
to the inner disk and has an inclination to the
line of sight in the range $50 < i < 90$
degrees. 

If there is a supermassive black hole
in Cygnus A we expect the gravitational influence
of the black hole to be reflected in the emission line kinematics
in the core region.
However, existing ground-based spectroscopic observations of the kinematics of the ionized 
gas on a kpc-scale in the 
ionization cones reveal a complex picture
(Simkin 1977, Tadhunter et al. 1994, Stockton, Ridgeway
\& Lilly 1994, Taylor et al. 2003). 
At radial distances between 1 and 3 arcseconds in the ionization cone
to the north west of the nucleus, the results of Stockton et al. (1994) 
show that the brightest emission lines are split by
$\sim$300 km s$^{-1}$, with the greatest splitting coincident with the
axis of the cone (approximately the radio axis). A fainter, but higher
velocity component --- moving at -1500 to -1880  km s$^{-1}$ in the rest
frame of Cygnus A --- is also detected 1.2 arcseconds from the nucleus, 0.5
arcseconds to the N of the radio axis (Tadhunter 
1991, Tadhunter et al. 1994). These line splittings and 
high velocity components are consistent with the effects of outflows induced
by the radio-emitting jets, or by winds driven either by the quasar nucleus
or a circum-nuclear starburst. The linear structures of the emission line
filaments in the NW cone, visible in high resolution H$\alpha$ emission line images (Jackson et al. 1998),
are also consistent with the effects of a large-scale outflow.

Within a radial distance of 1 arcsecond of the nucleus
the emission line kinematics appear to be more quiescent, with a 
velocity gradient across the nucleus in the N-S direction (total amplitude
$\sim$200  km s$^{-1}$), and a
rotation-curve-like pattern observed out to radial distances of 2 arcseconds
to both the north and south of the nucleus
(Simkin 1977, Stockton et al. 1994). However, Tadhunter et al.
(1994) have noted that, despite the apparently quiescent emission line
kinematics close to the nucleus, there is a peak in the [OIII] line width
coincident with the radio core. The overall rotation pattern in the N-S
direction is likely to be associated with the large scale disk that causes
the patchy dust obscuration visible in high resolution continuum
images of the core region (Jackson et al. 1998).

To summarise the important features of the
published results that are relevant to measuring
the mass distribution in the central regions of Cygnus A: 
\begin{itemize}
\item [-] the host galaxy Cygnus A harbours a quasar
nucleus and is therefore  expected to contain a supermassive
black hole that drives the quasar activity;
\item [-] the circumnuclear torus acts as a natural coronograph,
blocking of the direct light of the quasar nucleus along our
line of sight, and allowing the spatially resolved
gas kinematics to be measured close
to the nucleus;
\item [-] there is a high-surface-brightness emission line
structure close to the core which might be used as a kinematic
probe of the mass distribution;
\item[-] the geometry and inclination are well-constrained
by radio and optical observations;
\item [-] the existing low resolution observations provide
clear evidence for rotation in the central kpc, with the rotation
axis aligned close to the radio axis.
\end{itemize}
However, the existence of patchy dust obscuration across the
central regions, 
and presence of disturbed emission line kinematics in the NW cone, suggest
that some caution is required when interpreting the emission line kinematics
in the core region of this highly active galaxy.

\section{Observations and Reductions}
\subsection{Near-IR observations}

Near-IR spectroscopic observations of Cygnus A were taken on the night of
the 22/23 May 2000 using the NIRSPEC spectrograph in
grating mode on the Keck II
telescope (see McLean et al. 1998 for a description of NIRSPEC). Long-slit
spectra were taken along two position angles: PA180 and PA105. Figure
1 shows the slit positions marked on a 2.0$\mu$m NICMOS/HST image of Cygnus A
(see Tadhunter et al. 1999 for details).  In
order to facilitate good sky subtraction, the object was nodded between
two apertures along the slit in an ABBA pattern, with an integration
time of 400s for each of the four exposures. 

\begin{figure*} 
\epsfig{file=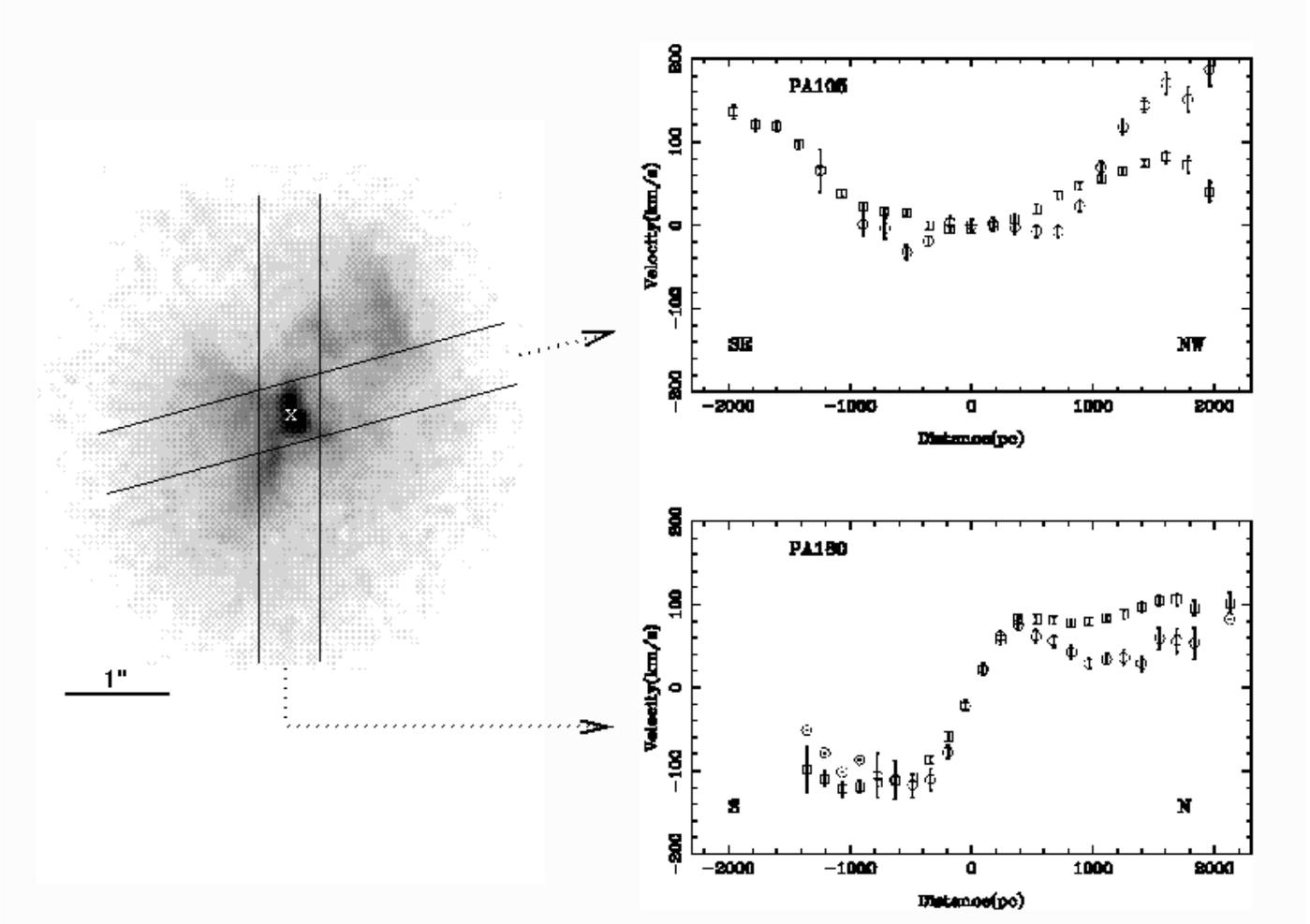,width=14.0cm}
\caption[]
{Infrared observations of Cygnus A. The left hand diagram shows the near-IR slit positions superimposed on the psf-subtracted 2$\mu$m
HST/NICMOS image (see Tadhunter et al. 1999 for details), while the right
hand plots show the radial velocity curves derived from the Keck/NIRSPEC
data (open circles: H$_2$ measurements; open squares: Pa$\alpha$ measurements).
In each case the zero point for the distance scale coincides with
the continuum centroid along the slit.}  
\end{figure*}
\begin{figure*} 
\epsfig{file=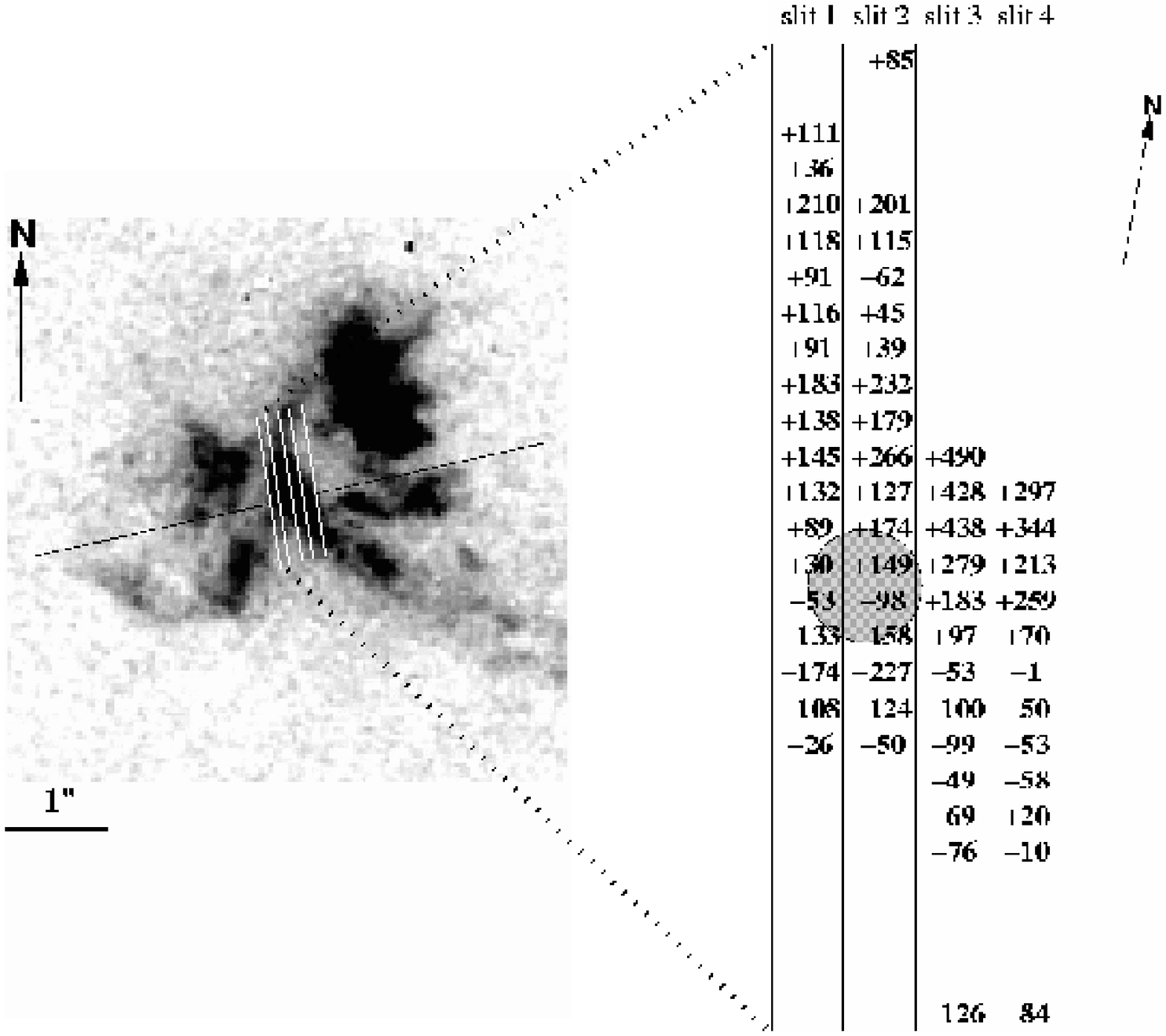,width=14.0cm}
\caption[]
{High resolution optical observations of Cygnus A. The left hand diagram shows the  STIS slit positions superimposed on the continuum subtracted 
HST/WFPC2 H$\alpha$ image of Jackson et al. (1998), while the right
hand diagram shows a map of the [OIII]$\lambda$5007 radial velocities
measured from the STIS data, with the estimated position of the nucleus
indicated by the centre of the shaded circle, and the uncertainty in
this position by the radius of the circle. The position angle of the radio
axis is indicated by the black line segment in the left-hand plot. 
}  
\end{figure*}

As a first step
in the data reduction, the spectra were rectified
using purpose-written IRAF routines, and arc-calibrated in IRAF using an
Argon/Neon arc lamp exposure taken at the time of the observations.
Following the manual removal of cosmic rays using the CLEAN routine in 
FIGARO, the separate sub-exposures for each PA were combined to produce
a sky-subtracted 2D image. Atmospheric absorption features were removed by dividing by
a high S/N spectrum of the standard star HD203856, which was observed
at a similar air mass. 
Use of the 0.57 arcsecond slit resulted in a spectroscopic resolution
of 9.9$\pm$0.5\AA\, ($\sim$ 150 km s$^{-1}$) with the observations covering the wavelength range
1.91 -- 2.3$\mu$m in the K-band. From measurements of stars along the
slit used for the Cygnus A observations we estimate a plate scale
of 0.178$\pm$0.004 arcseconds per pixel, and an effective seeing of 0.75$\pm$0.05
arcseconds (FWHM) for the observations. Measurements of
night-sky emission lines in the reduced frames demonstrate the wavelength
scale is accurate to within $\pm$0.5\AA\, ($\pm$8 km s$^{-1}$) along the
full length of the slit. 

%
 
\begin{figure*} 
\epsfig{file=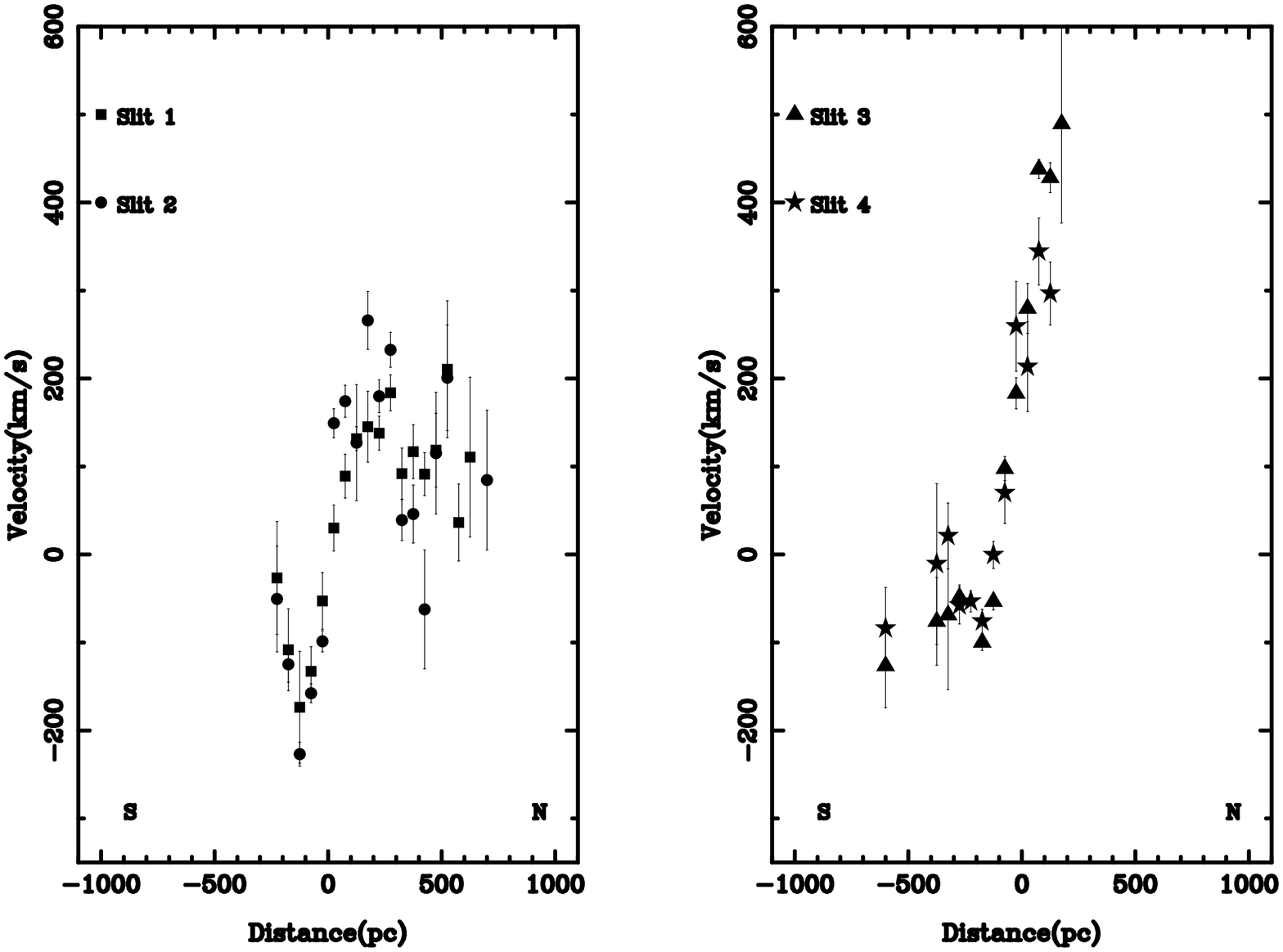,width=12cm,angle=270}
\caption[]
{[OIII]$\lambda$5007 radial velocities   
measured from the STIS data for the four slit positions: slits 1 and 2 (left);
slits 3 and 4 (right).}  
\end{figure*}
\begin{figure*} 
\epsfig{file=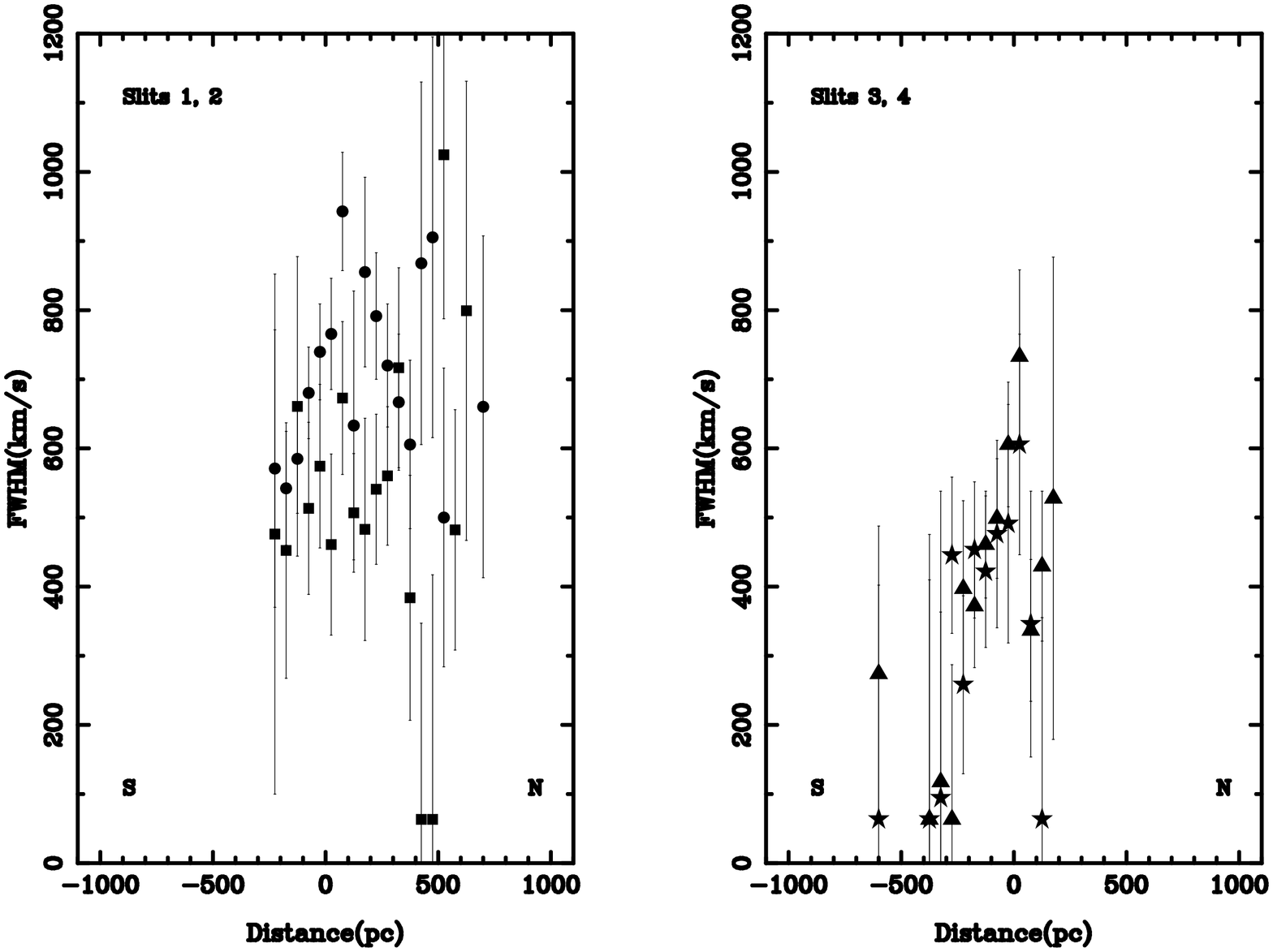,width=12cm,angle=270}
\caption[]
{[OIII]$\lambda$5007 instrumentally-corrected linewdiths    
measured from the STIS data for the four slit positions: slits 1 and 2
(left); slits 3 and 4 (right). The symbols are the same as in Figure 3.
}  
\end{figure*}


\subsection{Optical HST/STIS observations}

Our HST/STIS observations of Cygnus A comprise observations at
five parallel slit positions covering the high surface brightness emission
line structure in the core region of Cygnus A (Jackson et al. 1998). 
The position angle for the observations
was PA10 degrees (within  5 degrees of perpendicular to the radio axis).
A 0.1$\times$52 arcsecond slit was employed for these observations, and the slit
was offset successively by 0.1 arcseconds in the E-W direction
between the exposures. The slits were placed relative to the
core of Cygnus A by first using a peak-up routine to accurately
centre the slit on a nearby star, then performing an offset to the core
region of Cygnus A. For this purpose 
accurate core and offset star positions were determined
from our NICMOS images (Tadhunter et al. 1999). The positions of the slits
relative to the central emission line structures are shown in Figure 2.
Useful data were obtained in four slit positions.
We  label the slit furthest to the east slit 1 and the slit
furthest to the west slit 4;
slit 2 corresponds to the position of the
near-IR core source and contains the peak emission line flux. In order
to facilitate better correction for cosmic rays, hot pixels and other
cosmetic defects, the telescope was dithered by 10.5 pixels in the
direction parallel to the slit at each slit position, and
observations at each dither position were split into two separate
sub-exposures. The total
exposure time for each of the four useful positions was 2400s.

The first stages
of the data reduction were
carried out using the standard pipeline software. A major problem with 
STIS CCD data is the large number of hot pixels. In order to reduce
this problem we extracted the most up-to-date dark frame from the HST
archive to use in the reduction process. Purpose-written software
was then used to combine the two dither positions and thereby 
remove residual cosmic rays
and hot pixels.

Use of the F430L grating resulted
in a  linear dispersion of 2.75\AA\, per pixel, and
the instrumental width --- estimated from measurements of arc lines
in calibration lamp exposures --- is 6.25$\pm$0.4\AA\, (FWHM). 
On the
basis of the information given in the STIS instrument handbook
the estimated absolute uncertainty in the wavelength scale is 0.5 pixels
or 1.4\AA\, ($\sim$80 km s$^{-1}$)\footnote{The relative uncertainty
in the wavelength calibration between spectra taken in different
orbits but within the  5 orbit HST visit used for the Cygnus A observations is
likely to be smaller than this. Data for the individual slit position were
taken in different orbits and calibrated with arcs taken at the start of
each orbit. Based on the header information in the reduced files there is no evidence for a major change in the wavelength calibration
between orbits}.
The pixel scale in the spatial direction is 0.05 arcseconds per pixel, and
the estimated spatial resolution for the combined frames
is $\sim$0.1 arcseconds.

Both the STIS/HST and the NIRSPEC/Keck data were analysed using the
Starlink DIPSO package.
 


\section{Results}

\subsection{Kinematics based on near-IR lines}

Full analysis of the near-IR emission line spectrum is beyond the
scope of this paper, although we note that our Keck spectra
show spatially extended molecular (H$_2$), recombination (Pa$\alpha$)
and coronal ([SXI]$\lambda$1.92$\mu$m, [SiVI]$\lambda$1.96$\mu$m) lines along both slit position
angles, with a maximum spatial extent of $\pm$2 arcseconds ($\pm$2 kpc).
Here we concentrate on kinematic measurements derived from 
the bright Pa$\alpha$($\lambda$1.875$\mu$m) and H$_2$(1-0)S(1)($\lambda$2.121$\mu$m) lines, which are among the 
most extended. The radial velocity curves for the two slit position
angles are shown in Figure 1. In both of these plots
the measurements are based on single Gaussian fits to cores of the 
line profiles, even though it is clear that the broad wings to
Pa$\alpha$ --- detected within the seeing disk 
of the nucleus --- are not accounted for in 
such a simple model. In these plots the zero wavelength velocity
shift corresponds to a vacuum wavelength of 19805.0\AA\, ($z=0.05592$). 


Considering first the results for PA180 it is clear that the 
Pa$\alpha$ and H$_2$ lines show a smooth rotation curve, with
a steep gradient across the nucleus, and with the same sense
of rotation as measured for the optical emission lines
in both HST (see below) and ground-based spectra. The velocity
amplitude ($\pm$100 km s$^{-1}$) is also similar to that measured
on a kpc-scale in ground-based optical spectra, and outside
a radius of 0.2 arcseconds in our HST spectra.

In contrast, the spectra along PA105 (the radio axis) show no sign
of a steep velocity gradient across the position of the continuum
centroid, 
even though broad wings to the Pa$\alpha$ lines are detected close 
to the nucleus.

It is notable that significant differences in the emission line
kinematics between the Pa$\alpha$ and H$_2$ lines are present
along both PAs, although the two lines show a similar overall
pattern. The differences are clearest on the north side of the
nucleus along PA180, and to the north-west of the nucleus
along PA105 --- the region of line splitting detected
in ground-based optical spectra. These kinematic
differences suggest that the Pa$\alpha$ and H$_2$ lines may
sample different parts of the kpc-scale structure surrounding
the nucleus of Cygnus A. However, there is good agreement
between the kinematics of the two lines close
to the nucleus, out to a radial distance
of $\pm$0.5kpc. 

\subsection{[OIII] kinematics in the core of Cygnus A}

We present the results of single
Gaussian fits to [OIII]$\lambda\lambda$5007,4959 lines
detected in the STIS/HST data in Figures 2, 3 and 4.
A two dimensional map of the radial velocities is shown in Figure 2, 
while the variations in radial velocity and line width as
a function of slit position are plotted 
in Figures 3 and 4.  In these plots the zero
velocity shift corresponds to a vacuum
wavelength of 5288.5\AA\, (reshift:
$z = 0.05596$) --- we will discuss whether this zero redshift 
is appropriate in section
4.3 below. It is notable that the velocity
curves for all four slit positions show 
steep velocity gradients within a radius of $\pm$0.1 arcseconds
($\pm$100pc) of the assumed position of the nucleus. The velocity 
amplitude is -220 km s$^{-1}$ to the south of the nucleus, and
+420 km s$^{-1}$ to the north of the nucelus, and the steepest
gradient is measured in the slit position centred on the position
of the near-IR core source (slit 2). 
Outside a radial distance of 0.1 arcseconds,
the velocity curves measured for all four slit positions
appear to flatten out, and the velocity amplitudes ($\pm$100 km s$^{-1}$)
are consistent with those measured in ground-based long-slit spectroscopy and
the results from the near-IR spectroscopic results reported
in section 4.1. If all the velocity variations are
interpreted in terms of rotational motion, the sense of rotation 
on a $<$100pc scale appears to be the same as that on a 0.1 --- 1 kpc
scale.


However, despite showing clear evidence for rapid rotation across the nucleus,
not all the features of the STIS observations are consistent  with simple
rotation in  a planar, filled disk structure surrounding the nucleus which has
a rotation axis coincident with the large-scale radio axis. 
Most notably, within the central $\pm$0.1 arcseconds,
the velocity curves measured in slits 3 and   4  appear to
be displaced by -150 to -200 km s$^{-1}$ relative to the
velocity curves measured in slits 1 and 2 (see Figures 2 and 3). 
There is no clear
evidence for
such a displacement on a larger radial distances 
along the slit (although the data are
noisier on such scales). Also, the degree of displacement
is larger than can be accounted for by systematic errors in the 
absolute wavelength calibration. Therefore, these displacements
appear to be real.

As well as the radial velocity variations, we have also estimated the
widths (FWHM) of the [OIII] emission lines at all locations across the
core region. The line widths are
plotted against position in Figure 4. These have been
quadratically corrected for the instrumental profile
using the instrumental width determined from the arc exposures
(6.25$\pm$0.4\AA). The fact that the narrowest
line widths measured in the outer parts of the 
core region are close to the assumed instrumental width suggests that
this instrumental width is appropriate for our observations.

The striking feature of Figure 4 is that the lines are not only broad 
in the region closest to the nucleus (within a radial distance of
0.1 arcseconds), but remain broad ($400 < FWHM < 600$ km s$^{-1}$)
out to radial distances of 0.3 arcseconds on
either side of the nucleus. 
Although the broadest lines 
($600 < FWHM < 900$  km s$^{-1}$) are  detected in the pixels closest
to the nucleus, there is no evidence from these data
that an unresolved inner narrow
line region (INLR) contributes a large fraction of the flux in nuclear 
regions at these wavelengths.


\subsection{Comparison between near-IR and optical results}

Optical observations of the core region are potentially distorted
by extinction in the circum-nuclear dust lane. Indeed, there is
clear evidence for patchy dust obscuration from previous optical
high resolution HST images (Jackson et al. 1998). Such
obscuration could have the consequence that we see only
foreground material, and miss the line of
nodes of more rapidly rotating material buried deeper
in the dust lane. The near-IR spectroscopic observations should 
not be significantly affected by the dust obscuration
on this scale. Therefore, in this section we attempt to
assess the extent to which the dust extinction distorts the velocity 
profiles, by comparing various optical and infrared datasets.

The similarity between the shapes and amplitudes
of the large-scale IR (Pa$\alpha$ and H$_2$), the ground-based optical
([OIII], H$\alpha$: Stockton et al. 1994), and HST/STIS optical ([OIII]) rotation curves
suggests that, at least on a scale of 0.2 --- 2kpc, the optical
observations are not missing any high velocity, rapidly rotating 
components that are hidden by dust extinction in the 
core regions.

As well as the {\it shapes} of the spatially-resolved radial
velocity curves, it is also important to compare
the redshifts corresponding to the assumed zero velocities for
different datasets, in order to investigate
whether there are any systematic offsets in the {\it absolute}
velocities. Table 1 compares the various systemic redshifts that
have been obtained for Cygnus A over the last three decades. It is
immediately clear from this table that there remains some uncertainty about the systemic redshift. In particular, some of the recent studies based
on the optical/near-IR stellar absorption features give significantly lower values for
the redshift than the optical and infrared emission lines and
the HI 21cm radio absorption line. Otherwise,
the various emission line-based estimates are consistent with the low
redshift component of the  HI 21cm absorption line within the
errors.

Taken at face value, the differences between the emission and absorption line
estimates might be taken to 
imply that the entire circum-nuclear disk associated with the
dust lane is undergoing systematic motion relative to the stellar halo
of the host galaxy. The fact that the effect is seen in the IR emission lines
rules out the possibility that the true velocity distribution
is in fact symmetric  relative to the stellar halo and we are missing blueshifted wavelengths due to dust extinction. However, there also exists 
clear evidence
that the absorption line profiles are distorted due to infilling by
nearby emission lines (Stockton et al. 1994, Thornton et al. 1999). 
Therefore, while we cannot entirely rule out
the possibility that the disk/dust lane is moving relative to the stellar 
halo, it appears more plausible that the disk/dust lane is stationary 
with respect to the stellar halo, and the absorption line estimates
are systematically too low because of the emission line contamination.

\begin{table*}
\begin{tabular}{lll} \hline \hline
{\bf Type of Measurement} &{\bf Redshift} &{\bf Reference} \\
\hline
{\bf Ground-based absorption lines} & & \\
MgIb absorption, 5'' N of nucleus &0.0567$\pm$0.0016 &Sprinrad \& Stauffer (1982) \\
MbIb absorption in core &0.05562$\pm$0.00015 &Stockton, Ridgway \& Lilly (1994) \\
CaII triplet absorption in core &0.05544$\pm$0.0004 &Thornton et al. (1999)\\
\hline
{\bf Ground-based emission lines} & & \\
$[$OIII$]$, $[$NII$]$, H$\alpha$ radial velcity curves &0.05597$\pm$0.00015 &Stockton, Ridgway \& Lilly (1994) \\
Broad $[$OIII$]$ in core ($r < 0.5$ arcsec) &0.05606$\pm$0.00012 &Taylor et al. (2002) \\ 
Pa$\alpha$, H$_2$  radial velocity curves &0.05592$\pm$0.00005 &This paper \\
Broad Pa$\alpha$ in core ($r < 0.3$ arcsec) &0.05594$\pm$0.00005 & This paper 
\\ \hline
{\bf Radio observations of HI 21cm absorption line} & & \\
HI absorption against core (high velocity) &0.05663$\pm$0.00002 &Conway \& Blanco (1995) \\
HI absorption against core (low velocity) &0.05600$\pm$0.00002 &Conway \& Blanco
(1995)
\\ \hline
{\bf Space-based (HST/STIS) emission lines} & & \\
$[$OIII$]$ radial velocity curves ($r > 0.1$ arcsec) &0.05596$\pm$0.00024 
&This paper \\
$[$OIII$]$ (broad) spatially integrated ($r < 0.3$ arcsec) &0.05616$\pm$0.00024
&This paper \\
\end{tabular}
\caption[Redshifts]
{Summary of systemic redshift estimates for Cygnus A from this paper
and various literature sources. In the case of estimates based on
spatially resolved radial velocity curve information, the estimates represent
the redshifts required to make the putative rotation curves symmetric in velocity
about the nucleus.}
\end{table*}

As a final check that we have 
not missed high velocity components  at optical
wavelengths because of dust extinction, we can compare the STIS/HST [OIII]
and NIRMOS/Keck kinematics for the gas on an $r < 0.3$ arcsecond
scale. Although we do not resolve the Pa$\alpha$ and H$_2$ velocity
fields on this scale at the resolution of our Keck observations, 
we do detect broad wings to these lines within the seeing
disk close to the nucleus. If we assume that these broad wings
are associated with the unresolved  circum-nuclear regions, 
we can then make a direct comparison with the spatially-integrated
[OIII] emission line profile from our STIS/HST data. Figure   5  shows 
the best-fitting two Gaussian (broad+narrow)
line fit to the profile of Pa$\alpha$ in the nuclear regions.
The broad component to Pa$\alpha$ has a width of 750$\pm$50 km s$^{-1}$
(FWHM)\footnote{Note that what we label here as the ``broad'' component
cannot  represent transmitted or scattered light from a genuine, high density
broad line region (BLR) close to the central AGN, because this 
component is also detected in the forbidden lines.}. This compares with 850$\pm$50 km s$^{-1}$ obtained from
a single Gaussian fit to the profile of [OIII] integrated
over all four viable slit positions of our STIS/HST
observations out to a radius of $\pm$0.3
arcseconds from the spatial position of the core. The redshift
of the broad component measured in the near-IR data is also consistent
with that measured from the  
integrated [OIII] line profile in the STIS/HST data to 
within the errors. Overall, these results are consistent with
the idea that the broad component to Pa$\alpha$ samples the same velocity
structure as the [OIII] line in the circum-nuclear region, and that
the HST/STIS observations do not miss high velocity components, even
on the $r < 0.3$ arcsecond scale. They also suggest that the 
narrow emission line component --- detected in  Pa$\alpha$ ---
makes a negligible contribution to the STIS/HST
observations for $r < 0.3$ arcseconds. The fact that this latter component
is detected in the ground-based IR and optical observations is
likely to be a consequence of the wider slit used for the ground-based
observations and seeing disk spill-over from the bright extra-nuclear
emission line regions to the NW and SE. 

\begin{figure} 
\epsfig{file=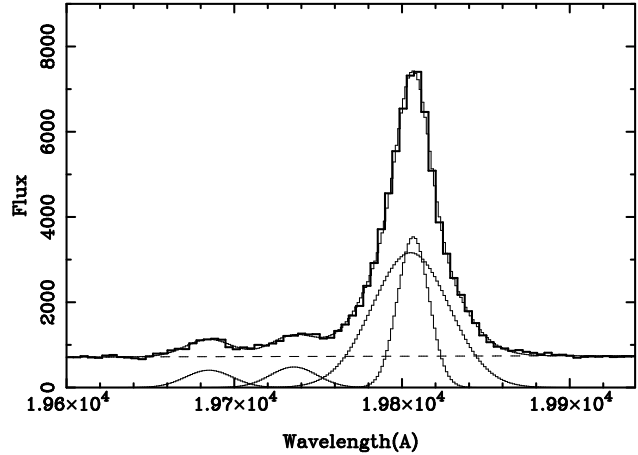,width=6cm,angle=270}
\caption
{Multiple Gaussian fit to the Paschen$\alpha$ profile in the nuclear regions of
Cygnus A, showing the broad and narrow components required to fit the 
profiles. The two components in the short wavelength wing of the profile are
HeII$\lambda$1.864$\mu$m and HeI$\lambda$1.869$\mu$m recombination
lines. This spectrum was extracted from the PA180 long-slit 
Keck data using a 0.72$\times$0.57 arcsecond aperture centred on the continuum centroid.}  
\label{figure:profile}
\end{figure}

\section{Discussion}


\subsection{The nature of the emission line kinematics}

A radial distances of 0.2 --- 1 kpc, the optical and IR emission
line kinematics are consistent with rotation about an axis that
is close to that of the large scale radio structure (see also Simkin
1977). At smaller radii the HST/STIS data also show some features that
are consistent with rotation in the same sense as the larger-scale gas,
but the velocity shifts measured between the  four  STIS slits are inconsistent
with simple rotation in a planar disk with a rotation
axis aligned along the radio axis. Possible explanations
for these shifts include the following.
\begin{enumerate}
\item{The rotation axis is significantly offset with respect to the
radio axis.} If the individual slit position angles were significantly
offset from the perpendicular to the rotation axis, this would lead
to apparent radial velocity shifts between the velocity
curves from different slit positions. This explanation would imply that
the rotation axis of the gas on a $r < 0.3$ arcsecond scale is 
substantially different
from that of the large scale disk/dust lane. 
\item{Non-circular motions distort the velocity field close to the nucleus.}
There is already clear evidence for radial flows in Cygnus A including
the line splitting measured at radial distances of 1 ---   3  kpc in
the NW cone (Stockton et al. 1994), and  the redshifted [OIII] emission line components
detected in polarized light (Ogle et al. 1997). If these
features are indeed due to radial flows, the flows are moving at
100 --- 500 km s$^{-1}$ relative to the nucleus.   
\end{enumerate}

Given that the most highly redshifted component to the NW is not far
from the radio axis 
(see Figure 2), the first of these possibilities would imply that, 
while the gas  in the large-scale dust lane is
rotating about an axis that is close to the radio axis, {\it some} of
the gas in the near-nuclear regions ($r < 0.3$ arcseconds) 
would be rotating about
an axis that is almost perpendicular to the radio axis. However, other
components in the near-nuclear regions would be following the rotation
pattern of the large scale disk structure. It is difficult to conceive
of a situation in which the two gas streams with perpendicular
rotation axes could survive on a reasonable timescale. In order
to avoid the intersection, the orbits of the gas streams would
require different radii, but the highly redshifted component 
to the NW 
has velocity amplitude which is greater than that  of the gas on a smaller
scale along the north and south 
edges of the bicone. Moreover, our modelling work shows that the implied mass
of the black hole ($> 10^{10} M_{\odot}$) would be inconsistent with
the results derived from the near-IR data (see section 5.2).
Overall, this
explanation appears to be untenable.

Therefore, the most plausible explanation for the large redshifts measured in 
slits 3 and   4  to the west of the nucleus is that we are observing a
component that is
not undertaking the pure circular rotation motion of the large-scale
disk/dust lane. We discuss the nature of these non-circular motions in
section 5.5.

\subsection{The mass of the MDO in Cygnus A}

In order to investigate the mass distribution 
we followed the method described in detail by 
Marconi et al. (2002). This consists of the two 
main steps:
\begin{enumerate}
\item The stellar contribution to the gravitational potential in
the nuclear region is obtained by assuming that the intrinsic geometry
of the stellar density distribution 
is an oblate spheroid with constant mass-to-light ratio and by comparing
the expected surface brightness distribution in the plane of the
sky with the observed one.
\item The gravitational potential given by the stellar mass distribution
obtained in (i) and a possible MDO are used to model the observed rotation
curves, assuming that the gas is in a thin, circularly rotating disk
in the principal plane of the stellar potential.
\end{enumerate}



The first step was to obtain the
surface brightness distribution of the starlight. However, 
a complication in the case of Cygnus A is that the near-nuclear regions are likely to be heavily extinguished at optical wavelengths by
the circum-nuclear dust lane (see Jackson et al. 1998). Therefore, we
used our near-IR observations taken with NICMOS/HST
to estimate the mass contribution of stars in the near-nuclear 
regions. The 1.6$\mu$m (F160W) image presented in Tadhunter et al. (1999) provides
a good compromise between decreased extinction (due to long wavelength)
and a relatively small contribution of the nuclear point source to the
flux close to the nucleus.
The measured  1.6$\mu$m light profile
is shown in Figure 6. Because of the potentially high
extinction due to the dust lane and the significant
flux contribution from near-nuclear illumination
cones (Tadhunter et al. 1999), 
in fitting the light profile, we have  only considered points 
outside a projected radius of $r> 0.3$ arcseconds.

\begin{figure} 
\epsfig{file=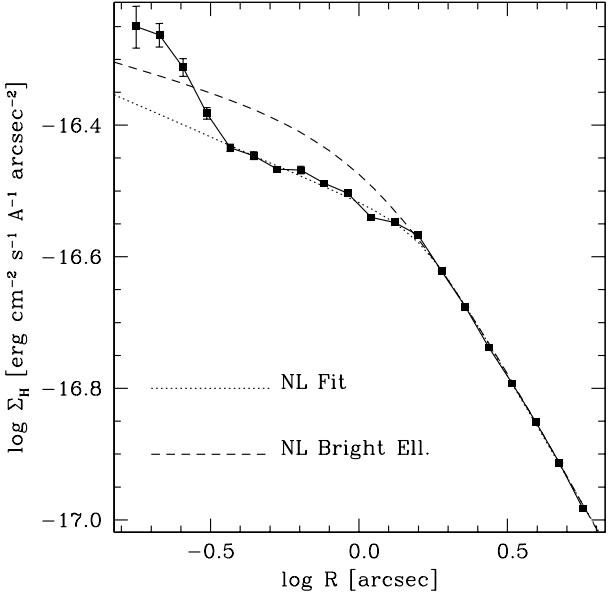,width=8cm}
\caption
{Fits to the H-band light profile in the inner regions of Cygnus A, 
based
on Nuker-law profiles (see text for details). The dotted line
shows a Nuker law fit to the measured points outside
a radius of 0.3 arcseconds, while the dashed line shows the
shape of the
Nuker law fit to the profile of the luminous
elliptical galaxy NGC4874 from Faber et al. 1997 (central
part only), 
normalised to the measured H-band 
flux of Cygnus A at a radius of 1.6 arcseconds. Note that
the peak in light profile at $log(R) < -0.5$ is due 
to contamination by the AGN.}  
\end{figure}

Given that the extinction may be significant even at
near-IR wavelengths, it is important to assess whether
the H-band radial light profile in
the central regions of Cygnus A 
is consistent with that expected
for a luminous early-type galaxy. Therefore we
fitted a ``Nuker'' law, which  Faber et al. (1997)
have shown to provide a good fit to the optical surface brightness
distributions of nearby elliptical galaxies. The Nuker law can be written as:
\begin{equation}
I(r) = I_b 2^{(\beta-\gamma)/\alpha}\left(\frac{r_b}{r}\right)^{\gamma}\left[1 + \left(\frac{r}{r_b}\right)^{\alpha}\right]^{(\gamma-\beta)/\alpha}
\end{equation}
where $r_b$ is the break radius. The dotted line in Figure  6 
shows the Nuker law fit to the
1.6$\mu$m (F160W) profile. This fit yields $r_b = 1.63$ arcsec (1.63 kpc),
$\alpha = 6.47$, $\beta = 0.774$, $\gamma = 0.197$. Although these values 
of $\alpha$, $\beta$ and $\gamma$ are within the range measured for
giant elliptical galaxies with similar absolute
magnitude to Cygnus A, the break radius is significantly
larger than measured for any of the giant elliptical galaxies
with core profiles in Faber et al. (1997). This may suggest that
the dust extinction has a significant effect on the near-nuclear
H-band light profile of Cygnus A. However,
the dashed line shows the Nuker law fit for the giant elliptical
galaxy NGC4874 -- which has a similar absolute magnitude to Cygnus A ---
normalised to the measured flux of Cygnus A at a radius of 1.6 arcsec
(at the outer edge of the dust lane visible in optical images
of Cygnus A). The fact that this dashed curve does not fall much
above the Nuker law fit to the 1.6$\mu$m profile suggests that the
effect of the dust lane extinction is relatively small --- at the $\sim$10\%
level.

Having demonstrated that the H-band light profile is 
not  heavily affected by extinction, the next step was to
derive the stellar density profile
assuming  a constant mass-to-light ratio and
oblate spheroidal geometry (step (i) above).
We assumed the following form for the stellar density
profile (see Marconi et al 2002):
\begin{equation}\label{eq:massdens}
\rho(m) = \rho(r_0)\left(\frac{m}{r_0}\right)^{-\delta} \left(1+\left(\frac{m}{r_0}\right)^2\right)^{-\epsilon}
\end{equation}
where $m$ is given by
\begin{equation}
m^2 = x^2+y^2+z^2/q^2
\end{equation}
and $xyz$ is a reference system with the $xy$ plane corresponding to the
principal plane of the potential. $q$ is the intrinsic axial ratio of the gravitational potential. The expected radial light profile was obtained by integrating the stellar luminosity density along the line of sight and convolving it with the psf. The $\rho(r_0)$, $r_0$, $\delta$ and
$\epsilon$ were then adjusted to give the best fit to the
data. As before, we  only used points at $r > 0.3$ for the fits.
The best fit -- shown as the solid line
in Figure  7 -- 
assumes that the gravitational potential is spherical
($q=1$). This is a reasonable approximation, given that the ellipticity of the 
isophotes
is around 0.2, i.e. their observed axial ratio is 0.8.
For an assumed mass-to-light ratio of $M/L_H = 1$  
we obtained the following fit
parameters: $r_0 = 1.41$ arcsec, $\delta = 0.0$, $\epsilon =0.946$,
$log(\rho(r_0)) = -0.225$ M$_{\odot}$ pc$^{-3}$. The case of a ``disk-like''
light distribution (Figure 6, dashed line), with an intrinsic axial ratio of 
$q=0.1$, gives $r_0 = 1.61$ arcsec, $\delta = 0.241$, $\epsilon =0.834$,
$log(\rho(r_0)) = 0.423$ M$_{\odot}$ pc$^{-3}$. It addition to fitting
the density law to the measured points (see Figure 7), we have 
also fitted it to the
Nuker profile given by the dashed line in Figure 6. This allows
the maximum likely contribution of stars to the mass in the
nuclear regions to be estimated.

\begin{figure} 
\epsfig{file=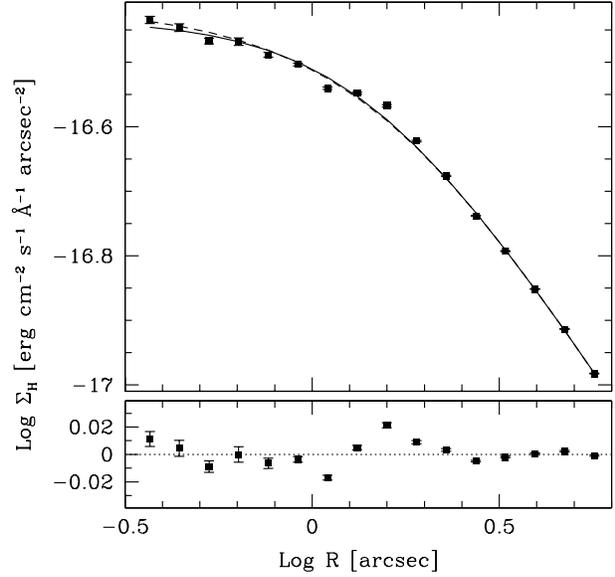,width=8cm}
\caption
{Fits to the H-band surface brightness profile assuming the stellar density
law given in the text (equation 2). The solid line assumes a spherical
density distribution, whereas the dashed line assumes a disk-like 
density distribution.}  
\end{figure}



Using the shape of the stellar density profile derived from
the surface brightness fits we then modelled the near-IR radial
velocity data, assuming there is a point-like massive dark
object (MDO)
in the nucleus of the galaxy, and that the inclination
of the rotation axis of the disk to the line of sight
is $i=55$ degrees. In adopting the latter inclination we assumed that
the emission line disk is perpendicular to both the jet and the illumination
cones axes (see section 2). This modelling takes
full account of the point spread function or seeing disk of the observations,
as well as the finite slit width (see Marconi et al. 2002 for full details).

Our models
require a functional form for the intrinsic line flux distribution. 
We find that,
after convolution with the psf and binning into pixels,
the line flux  distribution along the slit is well-fitted by 
two radially symmetric exponentials with scale radii of 0.39" and 0.48" respectively.
However, the line flux distribution can also be  adequately fitted with a combination
of two Gaussians, or a Gaussian and a constant.

Using the two exponential functional form for the line flux distribution we
fitted the radial velocity curves along PA180 and PA105 leaving
the MDO mass, the
mass-to-light ratio of the stellar populations
$\Upsilon$, the position of the nucleus along the slit, the
systemic redshift, and the PA of the line of nodes as free parameters. 
The best fitting model has: $M_{\small MDO} = 2.5\times 10^9$ M$_\odot$,
$\Upsilon=0.76$ and a position angle for the line of
nodes of  6 degrees. Note that the latter is close to the 15 degrees
expected if the disk is exactly perpendicular to the radio
axis; the systemic redshift
and position of the nucleus along the slit are also
consistent with the values
assumed in Figure 1 and Table 1. The results of the fitting are 
shown in Figure 8.

\begin{figure} 
\epsfig{file=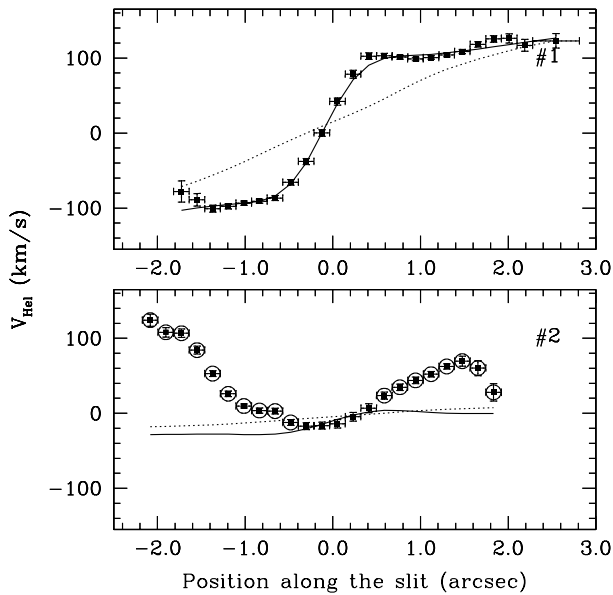,width=8cm}
\caption
{Fits to the radial velocity curves derived from measurements of
the near-IR Pa$\alpha$ line along PA180 (top) and PA105 (bottom).
The solid line shows the best-fitting model that includes an MDO, while
the dotted line shows the radial velocity curve that would be
predicted if there was no MDO and all the mass in the nuclear
region was due to the stars (from the fitted stellar density
profile). Both of the model curves takes into account
the line flux distribution and the psf.  The circled points are
likely to be affected by outflows in the cones and are therefore not included 
in the fits. N is to the right in the upper plot, and NW is to the right
in the lower plot.}  
\end{figure}

We attempted to estimate the errors on
$M_{\small MDO}$ and $\Upsilon$, based on a $\chi^2$ analysis
of a two parameter variation about
the best fit model. At the 3-$\sigma$ level we obtained:
$M_{\small MDO} = 2.5\pm0.4 \times10^{9}$ M$_\odot$ and  $\Upsilon = 0.76\pm0.2$.
This fitted $\Upsilon$ is consistent with the $M/L_H \sim 1$ predicted
by the spectral synthesis models for an old stellar population with
age $\sim 10$ Gyr (assuming a Scalo initial mass function).
We also explored the sensitivity of the fit to other parameters
and assumptions and found that the fit is not sensitive to: 
(a) the functional form assumed for the line flux distribution; (b)
whether we fit the PA180 data alone or both PA180 and PA105 together;
and (c) the assumed form of the stellar flux/density profile
(i.e. whether Nuker, spherical, or disk-like). We also found that, for
the stellar density profile derived from the H-band images, we
could not obtain an adequate fit to the data without a point-like MDO
at the centre of the Cygnus A (see the dotted line in
Figure 8). Perhaps the largest uncertainty in
estimating the mass of the MDO is the inclination: we  
assumed $i = 55$ degrees, but the inclination could fall in the
range $50 < i < 90$, based on the constraints provided by various 
optical and radio data (see section 2). 
Taking into account this uncertainty, we
have $M_{\small MDO} = 2.5\pm0.7\times10^{9}$ M$_\odot$.     
 
It is important to consider whether the model that provides the best fit
to the  near-IR data is also consistent with the higher spatial
resolution STIS data.
Having taken into account the different instrument setup
(psf, slit, pixel scale) for the STIS observations, in Figure 9 we compare the radial velocity measurements 
derived from the four viable STIS slit
positions with the predictions of the best fit near-IR model. It is clear that,
while the model provides a good fit to the slit 1 and slit 2 data, the slit 3
and   4  data show a large redshift with respect to the model. Therefore,
assuming that the Pa$\alpha$ and [OIII] sample the gas kinematics in the
same disk with the same geometry, this reinforces the conclusion that
the [OIII] measurements along slits 3 and   4  are affected by non-circular
motions (section 5.1).

Figure 10 shows a more detailed comparison between  the
slit 1 and 2 results and model
predictions for  MDO masses in the range  
$1\times10^9 < M_{MDO} < 6\times10^9 M_{\odot}$. 
It is clear from this that the data are
bracketed by the predictions of the 1$\times$10$^9$ and 
6$\times$10$^9$ M$_{\odot}$ models, and are best fitted by the 
2.5$\times$10$^9$ M$_{\odot}$ model.

\begin{figure} 
\epsfig{file=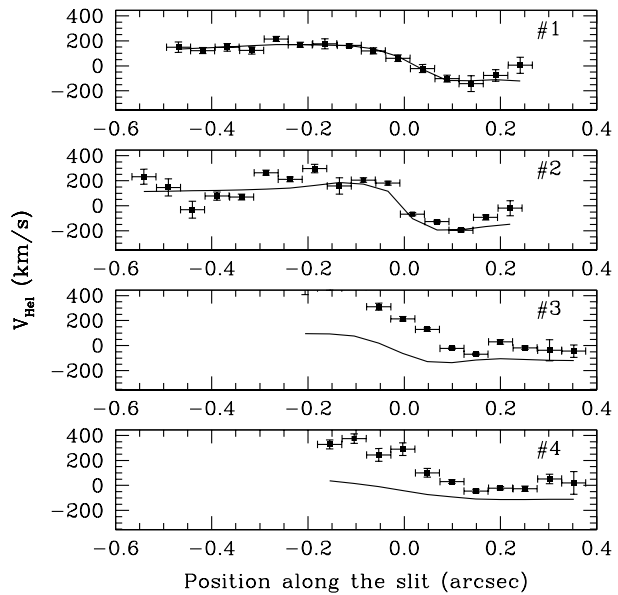,width=8cm}
\caption
{Comparison between the radial velocities measured from the STIS data
along four slit positions and the model predictions based fits
to the near-IR data. In these plots N is to the left.}  
\end{figure}
\begin{figure} 
\epsfig{file=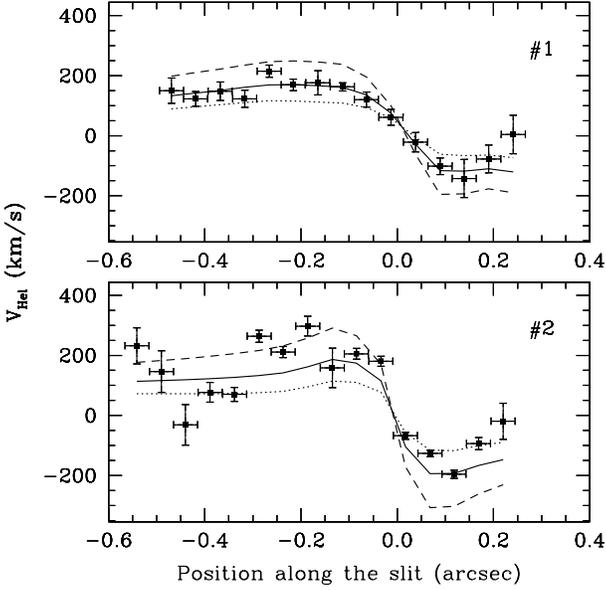,width=8cm}
\caption
{Detailed comparison between the data and the models for slits
1 and 2 for different MDO masses. The three model curves are
for MDO masses of 2.5$\times$10$^9$ M$_{\odot}$ (solid),
1$\times$10$^9$ M$_{\odot}$ (dashed) and  6$\times$10$^9$ M$_{\odot}$ (dotted). Note that the fit of the 2.5$\times$10$^9$ M$_{\odot}$ model could
be improved by moving the measured radial velocities for slit 2 
systematically down by 50 km s$^{-1}$
(this is within the estimated systematic uncertainty of the wavelength
calibration). In these plots N is to the left.}  
\end{figure}

We also fitted all   four  slit positions of the
STIS data allowing both the PA of the line
of nodes and the mass of the black hole to vary. In this case we
found that the best fit to the [OIII] data for
all four slit positions modelled together was obtained for an MDO of mass  $M_{MDO} > $10$^{10}$ M$_\odot$ with  a position
angle of the line of nodes of -45 degrees. Clearly this model is inconsistent with the
near-IR data, and also fails to fit the data from all four STIS slit positions
adequately. 

The fact that the model based on the fit to the near-IR data also
provides an excellent fit to the slit 1 and 2 
STIS/HST data suggests that extinction effects are
not causing us to miss
components of the disk with high projected radial velocities at
optical wavelengths, thus strengthening the conclusions
of section 4.3. Moreover, the steepness of the velocity gradient
across the nucleus along slit 2 --- the slit coincident with the
infrared nucleus --- demonstrates that the mass associated with the
MDO must be confined to a radial distance from
the kinematic centre of $r < 50$pc (corresponding to half the STIS
spatial resolution i.e. 0.05 arcseconds).

\subsection{Line width distributions}

Previous studies provide evidence for an increase
in line width towards the nucleus of Cygnus A (Tadhunter et al. 1994). This
increase is also seen in the near-IR and optical data presented
in this paper. In terms of
understanding of the NLR kinematics in radio galaxies,
it is important to consider the extent to which the large line
widths are due to unresolved rotation about the central MDO, and the
extent to which they are intrinsic.

The kinematic models presented
above are based entirely on fits to the radial velocity curves.
In order to determine the line width distributions from the
best fit model we have assumed the line profile at each location
in the rotating disk can be modelled with a Gaussian with an intrinsic
$\sigma$ that remains constant across the disk. The values of $\sigma$
were chosen to provide the best fit to the line widths in the outer parts of
the disk. Figures  11 and 12 compare the predicted line widths, based on the
model that fits the radial velocity curves best, with the
line widths measured from the data for the
near-IR and optical data respectively. For the near-IR data an intrinsic
$\sigma = 110$ km s$^{-1}$ was assumed, whereas for the optical data
$\sigma = 220$ km s$^{-1}$ was assumed.

\begin{figure} 
\epsfig{file=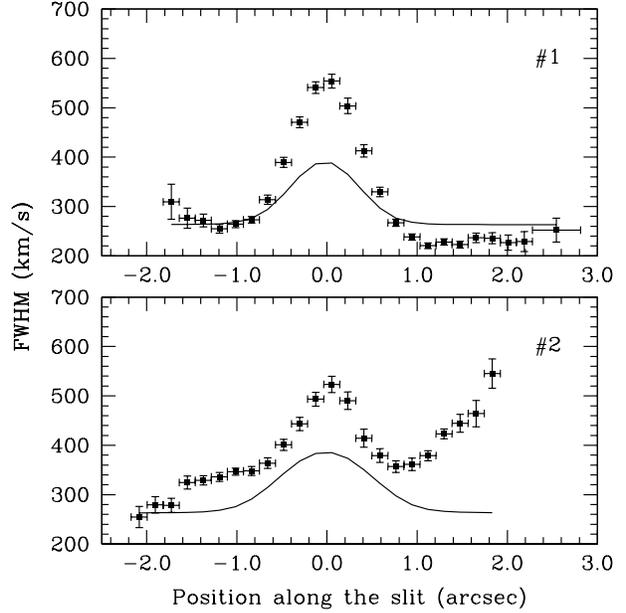,width=8cm}
\caption
{Pa$\alpha$ line widths measured along PA180 (top)
and PA105 (bottom) compared with
model predictions based on fits to the near-IR radial velocity curves.
The model predictions assume that each location in the disk has an intrinsic Gaussian
line profile with $\sigma = 110$ km s$^{-1}$. In this case the measured line widths have not been corrected for the instrumental profile, but the
models are convolved with the width of the slit in velocity space
(see Marconi et al. 2002). In the upper plot N is to the right,
while in the lower plot NW is to the right.}
\end{figure}

\begin{figure} 
\epsfig{file=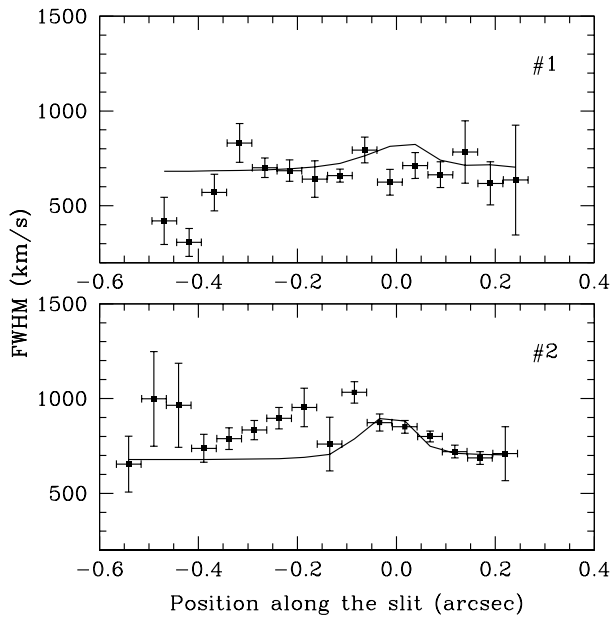,width=8cm}
\caption
{[OIII] line widths compared with
model prediction based on fits to the near-IR radial velocity curves.
The model predictions assume that each location in the disk has an intrinsic Gaussian
line profile with $\sigma = 220$ km s$^{-1}$. In this case the measured line widths have not been corrected for the instrumental profile, but the
models are convolved with the width of the slit in velocity space
(see Marconi et al. 2002). In these plots N is to the left.}
\end{figure}

 On the basis of this comparison we note the following.
\begin{itemize}
\item {\bf Unresolved rotation.} The large line widths cannot be solely due to unresolved rotation; there
is a large intrinsic line width at all locations in inner disk ($r < 0.4$
arcseconds).
\item {\bf Gradients in the intrinsic line width.}
Based on the failure to fit the large line widths measured close
to the nucleus in the near-IR data ($\sigma = 110$ km s$^{-1}$), and also the larger intrinsic linewidth
required for the optical data closer to the
nucleus ($\sigma = 220$ km s$^{-1}$), the intrisic linewidth must rise between the outer and inner parts of the circum-nuclear gas disk.
\item {\bf Larger scales ($r > 0.5$ arcseconds).} For the near-IR data, the fact that significantly larger linewidths 
are measured at $r > 0.5$ arcseconds along PA105 than on similar spatial
scales along PA180, is likely to be a consequence of unresolved line
splitting in the north west and south east cones, as detected 
in the optical data of Stockton et al. (1994) 
(Note: our measured line widths are based on single Gaussian fits to the line cores).
\end{itemize}

The cause of the large intrisic line widths close to
the nucleus ($r < 0.5$ arcseconds) is currently uncertain. One
possibility is that it is due to 
unresolved line splitting resulting from radial
outflows in the near-nuclear regions. In this case, the flux in the redshifted
component emitted by far side of the cone/disk would be required to have
a similar flux to theblueshifted component emitted by the near side 
of the cone/disk, in order to avoid the large line shifts observed
in slits 3 and   4 (see section 5.1).

Alternatively, the large line widths may be a consequence
of a large, activity-induced turbulence at each
location of the disk. For example, such 
turbulence  could be generated by instabilities in the boundary
layers between the warm gas in the walls of the hollowed out bi-cone
and the outflowing hot wind and jets.

Finally, we must also consider the possibility that the large line widths
are gravitationally-induced. For example, if the emission line structure 
comprises a collisionless system of cloudlets, it might be partially
supported against gravity by the random motions of clouds on
non-circular orbits 
(the so-called ``asymmetric drift''). In this case, given that 
intrisic 
Gaussian velocity width ($\sigma$) is comparable with
the bulk circular velocity ($v_c$) throughout the inner disk 
($\sigma/v_c \leq 1$), the 
MDO mass we estimated above on the basis
of the bulk circular motions alone would underestimate the true mass of
the MDO. We note that the 
asymmetric drift is a controversial issue
for gas dynamical estimates of black hole masses
in general (e.g. van der Marel \& van den Bosch 1998,
Verdoes Kleijn et al. 2000, Barth et et al. 2001, Verdoes
Kleijn et al. 2002). An argument against gravity-dominated line
widths in the case of Cygnus A is that the lines are broad throughout
the near-nuclear disk in the STIS/HST data, whereas they would be
expected to be more strongly peaked towards the nucleus
in the gravity-dominated case. 

Overall, given the extreme level of nuclear activity
present in this source,  the
large line widths measured across the core of Cygnus A are  
most likely be a consequence of local turbulence induced 
by the activity. In this case
the  circular velocities provide a true measure the MDO mass.

\subsection{Contributions to the mass in the near-nuclear regions
of Cygnus A}

Based on our analysis of both the near-IR and
optical data we have found evidence for $M_{MDO} = 2.5\pm0.7\times10^9$ M$_\odot$
concentrated within a radial distance of 50pc from the 
dynamical centre of Cygnus A. This corresponds
to a mean mass density of $4.7\pm1.3\times10^3$ M$_\odot$ pc$^{-3}$. 
By itself this large central mass density does not 
imply that the MDO is a black hole.
We now consider the possible contributions to the central
mass.

The nuclei of radio galaxies are hosted by a massive stellar
bulges. Therefore it is important to consider the contribution of stars to
the mass in the central regions. Unfortunately, in the case of Cygnus A the 
stellar contribution cannot be 
measured directly  because of the extinction
due to the kpc-scale dust lane
and circum-nuclear torus. Rather, we are forced to rely on 
extrapolations of the stellar mass density profile derived from the H-band 
imaging data, and comparisons with nearby inactive giant elliptical galaxies of
similar absolute brightness. The extrapolation of the
stellar density profile derived fitting the maximal Nuker-law
profile (dashed line Figure  6) 
gives a mean stellar mass density of 
$\rho_s = 10.0$ M$_\odot$ pc$^{-3}$ within
the central 50pc of Cygnus A. In comparison, the nine galaxies from Faber
et al. (1997) with absolute magnitudes similar to Cygnus A
($M_v < -22.0$) have mean central mass densities within a
similar radius in the range $23 < \rho_s <  160$  M$_\odot$ pc$^{-3}$.
Therefore the mass density measured from the
emission line kinematics in the core of Cygnus A is almost three
orders of magnitude larger than the stellar mass density estimated
from the H-band flux profile. It is
also two orders of magnitude larger than typical mean stellar mass
density estimated for giant elliptical galaxies of similar absolute
brightness. On this basis we conclude that the large central mass in Cygnus A 
is unlikely to comprise
{\it normal} stellar populations similar to those found in the cores of giant elliptical
galaxies. 

However, it is more difficult to rule out the idea that a compact star cluster
close to the nucleus makes a significant 
contribution to the central mass. On the basis of the presence of the central obscuring torus and kpc-scale
dust lane, it is clear that the central regions of
of Cygnus A are gas rich. There is also direct evidence for 
recent star formation in the outer parts of the kpc-scale disk from
optical HST imaging observations (Jackson et al. 1998, Fosbury et al. 1999). Therefore 
it remains a possibility that a massive star cluster
has formed recently the high density central regions of Cygnus A, and that
this nuclear star cluster contributes to the large mass density within
a radius of 50pc. 
Using
the results of isochrone  spectral synthesis models we
find that the optical luminosity
of a  $3\times10^9$ M$_{\odot}$ cluster would not be competitive with that
of the quasar nucleus in Cygnus A ($M_v \sim -23$) unless
the cluster was very young and the starburst instantaneous
(the Bruzual and Charlot 1996 models predict
$M_v > -20$ for age $>$100 Myr and starburst mass 
$\sim 3\times10^9$ M$_{\odot}$). Therefore, it may be relatively
difficult to detect such a star cluster, especially if it is hidden
from our direct view at optical and infrared wavelengths
by the material in the torus.

We must also consider whether warm or cold gas  
can make a significant contribution to the total mass density 
close to the nucleus.
On this scale it is likely that the dominant gas mass is contributed by the
circum-nuclear torus that obscures the quasar nucleus from our
direct view at optical wavelengths. To estimate the mass of the
torus we assume that it has a cylindrical geometry. In
this case, for the estimated
opening half-angle of the ionization cones ($\theta_{1/2}$), 
the total thickness of the torus is $t = 2 r_i \tan(90-\theta_{1/2})$, 
where $r_i$ is the radius
of its inner face. If we make the futher assumption that
the X-ray absorbing column ($N_{\small X}$) is representative
of the mean column depth through the torus in the direction perpendicular
to the torus axis, the mass of the torus is given by:
\begin{equation}
M_{torus} = 2 \pi r_i(r_i + r_o) \tan(90-\theta_{1/2}) N_{\small X} m_p  
\end{equation}
where $m_p$ is the mass of a proton, and $r_o$ is the outer
radius of the torus. For a cone opening half
angle of $\theta_{1/2} = 60$ degrees (Tadhunter et al. 1994, Jackson et al. 1998),
an inner radius of the torus of $3.1\times10^{19}$ cm (10pc) --- (see discussion in Blanco et al. 1995, 
Maloney et al. 1995), an outer radius equal to the
smallest radius sampled by our STIS data
(i.e. $r_o = 50$pc), and an X-ray absorbing column of $N_{\small X} = 2.3\times10^{23}$
cm$^{-2}$ (Young et al. 2002), 
we derive $M_{torus} = 3.9\times10^6$ M$_\odot$ --- almost three orders of
magnitude smaller than required by our dynamical estimate of the
mass within 50pc. The torus would be more massive if the measured X-ray column
substantially underestimates the mean column density of the gas of the torus.
For example, our line of sight might pass through the low density
outer layers
in the torus, or the material in the
torus might be clumpy and our line of sight might
miss the higher density clumps. However, in order for the torus to contribute most of
the mass within 50pc, the required column density in the higher density
regions would be 
several orders of magnitude larger than the absorbing
column sampled along our line of sight. 

In view of the possible --- if not likely --- contribution of stars
and gas to the mass in the nuclear regions, 
strictly the mass we have measured within
50pc is an upper limit on the mass of the nuclear black hole. On the
other hand, if the near-nuclear disk is partially supported
by the non-circular, gravitationally-induced motions of cloudlets (see section 5.3),  the 
mass estimate based on the bulk circular velocities
alone (section 5.2) underestimates the total mass within 50pc. In 
the following we will 
assume that the mass within 50pc, as estimated from the bulk
circular velocities, is an accurate reflection
of the mass of the nuclear black hole.

\subsection{Black hole mass and host galaxy
properties}

With an estimate of the black hole mass in Cygnus A we are now in a position to
discuss the links between the black hole and host galaxy
properties. Some basic data for 
Cygnus A, its AGNand its host galaxy  are listed in Table 2.

\begin{table*}
\begin{tabular}{lllll} \hline \hline
{\bf Property} &{\bf M87} &References &{\bf Cygnus A} &References\\
\hline
{\bf Host Galaxy and Environment} & & & & \\
Absolute magnitude ($M_R$) &-23.16 &[2] & -23.34 &[3]\\
Environment &Cluster & &Cluster &\\
\hline
{\bf Radio Source and AGN} & & & &\\
Radio morphology  &FRI & &FRII & \\
Radio power at 6cm ($P_{6cm}$) &$2.2\times10^{24}$ (W Hz$^{-1}$) &[4] &$5.4\times10^{27}$ (W Hz$^{-1}$) &[4]\\
Black hole mass ($M_{bh}$) &$3.2\pm0.9\times10^{9}$  M$_{\odot}$ &[5] &$2.5\pm0.7\times10^{9}$ M$_{\odot}$ &[1]\\
AGN Bolometric luminosity &$< 10^{43}$ (erg s$^{-1}$) &[6] &$0.5$ -- $2.0\times10^{46}$ (erg s$^{-1}$) &[1] \\ 
Eddington luminosity ($L_E$) &$4.0\times10^{47}$ (erg s$^{-1}$) &[1] &$3.3\times10^{47}$ (erg s$^{-1}$) &[1]\\
Eddington ratio ($L_{BOL}/L_E$) &$<2.5\times10^{-5}$ &[6] &($1.5$ -- $6.0$)$\times10^{-2}$ &[1] \\
Mass accretion rate (\.{m}) &$3\times10^{-3}$ &[6] &($1.5$ -- $6.0$)$\times10^{-2}$ &[1] \\ 
\end{tabular}
\caption[Redshifts]
{Summary of basic properties of the host galaxy, radio source and AGN
of Cygnus A. 
The mass accretion rate \.{m} is expressed as a fraction
of the mass accretion rate required to produce the
Eddington luminosity (\.{m}$_E$), and 
assumes a constant radiative efficiency factor. References:
1. This paper; 2. Faber et al. (1997); 3. Carilli et al. (1989); 4. 
Kellerman et al. 
(1969); 5. Marconi et al. (1997); 6. Reynolds et al. (1996);
7. Owen et al. (1997). Note that the
absolute R-band magnitude for M87 has been calculated from the 
absolute V-band magnitude of Faber et al. (1997) assuming that M87
is at a distance of 15.3 Mpc and (V-R) colour index of 0.6; 
the absolute R-band magnitude
of Cygnus A has been calculated from the R-band photometry of Carilli et al.
(1989) assuming a Galactic extinction in the R-band of 1.02 magnitudes. 
}
\end{table*}

First we note  that the estimated mass of the black hole and
R-band absolute luminosity of the host galaxy of Cygnus A are consistent
with the correlations between black hole mass and bulge luminosity
deduced for normal (non-active) galaxies. Based on the measured absolute
R-band luminosity of the host galaxy, the relationship between black hole mass
and bulge liuminosity derived by McClure \& Dunlop (2002) for inactive
elliptical galaxies predicts a black hole mass of $1.6\times10^{9}$ M$_{\odot}$ for Cygnus A. Given the uncertainties, this is consistent with
our dynamical estimate of the mass of the black hole in Cygnus A. 
Therefore, the mass of the black hole is about what one would expect on the basis of its stellar bulge luminosity; the extraordinary level of activity in Cygnus A cannot be explained in terms of
accretion onto an unusually massive black hole. 

Unfortunately, because of the potential contamination
of its stellar absorption lines by emission lines,
it is less certain whether
Cygnus A is consistent with black hole mass versus
stellar velocity dispersion correlation deduced for non-active
galaxies (Gebhart et al. 2000, Ferrarese \& Merritt 2000, Tremaine et al.
2002). 
Using
the stellar velocity dispersion measured by Thornton et al. (1999: $\sigma
= 290\pm70$ km s$^{-1}$) along
with the best fit $M_{bh}$ vs. $\sigma$ relationship from 
Tremaine et al. (2002), the predicted black hole mass is
$0.6^{+0.8}_{-0.4}\times10^{9}$ M$_{\odot}$. This is consistent
with, but somewhat lower than,
our gas dynamical mass estimate. However, as discussed in
section 4.3, the results of
Thornton et al. are likely to be affected by the infilling
of the stellar absoption lines by emission lines, and 
such infilling will lead to spuriously
low $\sigma$ and $M_{bh}$.

It is also interesting to compare our 
dynamical estimate for the mass of the black hole in
Cygnus A with the virial mass estimate derived from the properties
of the broad permitted lines detected
in scattered light at both UV and optical wavelengths (Antonucci, Hurt \& Kinney 1994, Ogle et al. 1997). Following Kaspi et al. (2000), the radius of the broad line
region (BLR) in Cygnus A is $\sim$180 lt-days for $L_{bol} = 10^{46}$ erg
s$^{-1}$\footnote{We assume here that $L_{bol} \sim 9\lambda L_{5100}$ erg
s$^{-1}$ (see Kaspi et al. 2000).}. Two broad line FWHM estimates are available in the literature: 
H$\alpha$(FWHM)$\sim$26,000 km s$^{-1}$ (Ogle et al. 1997) and 
MgII(FWHM)$\sim$7,500 km s$^{-1}$ (Antonucci et al. 1994).
However, 
we regard the larger H$\alpha$ linewidth derived from the spectropolarimetry observations
of Ogle te al. (1997) as the more reliable, since it is based on two independent sets
of spectropolarimetry data taken on either side of the nucleus; the
larger linewidth estimate is also consistent with the FWHM measured in
some broad line radio galaxies (BLRG) of comparable AGN brightness to Cygnus A
(e.g. Osterbrock, Koski \& Phillips 1976, Corbett et al. 2000).
Substituting the H$\alpha$ linewidth into the virial equation of Kaspi et al. (2000) gives $M_{bh} = 1.75\times10^{10}$ M$_{\odot}$. 
The fact that this is larger than our dynamical mass
estimate implies that there is
a non-gravitational (e.g. outflow) component to the BLR kinematics in nearby radio-loud AGN.
Such non-gravitational kinematics are consistent with recent spectropolarimetry results   
for BLRG (Corbett et al. 2000); they also help to explain why, in the Seyfert 1 sample
of McClure et al. (2001), the BLRG 3C390.3 has an unusually large virial black
hole mass for its host luminosity. Clearly, some caution is required when using BLR
properties  to estimate the masses of
the black holes in radio-loud AGN. 

\subsection{Implications for the nature of the AGN in Cygnus A}

In terms of understanding the nature of
the AGN in Cygnus~A it is interesting to consider the accretion rate required to produce the
observed level of nuclear activity. The estimate of the bolometric luminosity of the AGN in Table 2 was derived from the measured X-ray luminosity of the core using 
the  spectral energy distributions of radio-loud quasars presented by Elvis et al. (1994). This luminosity implies that that the mass accretion rate is a few percent
of that required to produce the Eddington luminosity
(i.e. $1.5\times10^{-2} < $ \.{m} $<  6.0\times10^{-2}$), assuming a constant efficiency factor.

For comparison we also include data for the nearby low-luminosity, FRI radio 
galaxy -- M87 -- in Table 2. The comparison with M87 is 
particularly interesting, because the mass deduced for the black hole in M87 is
similar to that determined for Cygnus A; its absolute magnitude and environment
are also similar. What distinguishes M87 from Cygnus A is that its AGN is
two to three orders of magnitude less luminous overall, and its radio jets
and lobes are three orders of magnitude less powerful. A further difference is that, whereas a broad-line quasar nucleus has been detected in spectropolarimetric observations of Cygnus A (Ogle et al. 1997), no such broad line nucleus has so far been detected in M87.

Given the similarity in their black hole masses, it is likely that the 
differences between M87 and Cygnus A are related to some
property other than black hole mass. Most plausibly this property is the mass accretion rate, which is likely to influence the 
type of accretion and the
radiative efficiency of the accretion process (e.g.  Reynolds et al. 1996). 
The overall properties of the AGN in M87 can be modelled in terms of a low efficiency advective accretion flow if the black hole 
is accreting at the Bondi accretion rate (Reynolds et al. 1996), whereas those of Cygnus A are entirely consistent with accretion via a standard, geometrically
thin accretion disk at much higher efficiency. In this
context it is interesting 
that the Bondi accretion rate deduced for M87
is only an order of magnitude less than the accretion rate deduced for
Cygnus A. This suggests that, if the mode of accretion (e.g. whether
advective, or standard accretion disk) is dictated mainly by the
mass accretion, the switch between the two modes occurs 
over a relatively narrow range of mass accretion rate: somewhere 
in the range $10^{-3} <$ \.{m} $< 10^{-2}$.

\subsection{Non-circular motions in the NW cone}

An interesting aspect of the STIS/HST observations 
is that a component of gas in the NW cone (slits 3 and 4) does
not fit in with the overall rotation pattern deduced for
the emission like gas from the other STIS data and
the near-IR observations. We now
consider the nature of these non-circular motions.

One possibility is that the non-circular motions are associated
with a cloud that is falling through the NW cone on a highly elliptical, near-radial path. For example, the cloud could represent
material that has condensed
out of the large-scale cooling flow but has not yet settled into
the circum-nuclear dust lane. This possibility is difficult to rule out, given
that warm/cool material could be continually falling into the nuclear
regions from the large scale cooling flow that has been deduced from
X-ray observations.
Even in the case of the gas originating in a galaxy merger
or encounter, it is possible that clouds will fall into the nuclear regions
on near radial orbits until all the gas finally relaxes into a stable 
configuration. 
It is also notable that a HI 21cm absorption line
feature is detected against the radio core that is redshifted
relative to the galaxy rest frame ($\Delta V = +179\pm10$ km s$^{-1}$:
Conway \& Blanco 1995, see Table 1) --- a clear sign that at least some
material is falling in towards the nucleus.
However, in the context of this explanation,
it seems remarkable that the redshifted component detected in the
STIS data is
observed so close to the axis of the cones and the large-scale radio
jets, and also that it appears to show evidence for differential
``rotation'' in the same sense as the gas closer to the nucleus
and the large-scale dust lane.  

Alternatively, the redshifted component
may represent line emission from gas that is undergoing a systematic radial
outflow from the nucleus. Indirect evidence for outflows in the near-nuclear
regions of Cygnus A 
is provided by the hollowing out of the cones apparent in
high resolution near-IR HST images of the core (Tadhunter et al.
1999). Several independent lines of evidence suggest that the cone
axes are tilted relative to the plane of the sky and that the NW
cone is in the foreground and the SE cone is in the background
of the circum-nuclear dust lane. In this case, in order to produce
a NW redshifted component with an outflow, the material must be flowing away
on the far side of the NW cone. Then, given that the NW cone
is tilted towards the observer, the de-projected radial velocities
of the outflow would be much larger than measured. Note that, for us to
observe significant line emission from the far side of the cone, the
extinction due to dust in the cone must be relatively small. 
It may also be of concern that there is no evidence close to
the nucleus for a blueshifted component associated with outflowing
material on the near side of the cone (expected to be less highly extinguished).

The geometrical constraints associated with the outflow model are less stringent
if we are observing line radiation scattered by an outflowing wind, 
rather than direct line radiation from the outflow itself. In this 
case, provided that
the axis of the cone is not too close to our line of sight, light scattered
from outflowing material filling the NW cone will produce a redshifted feature.
The scattering outflow idea is supported by the fact
that polarized, redshifted [OIII] lines have been detected in spectropolarimetry
observations of Cygnus A (Ogle et al. 1997); at the position of the 
nucleus the polarized lines are redshifted by $\sim480\pm80$ km s$^{-1}$
(van Bemmel 2002) --- close
to the redshift of the component detected to the NW of the nucleus in our
STIS data. The major uncertainty with this explanation is whether sufficient
scattering dust could survive in the hostile environment of
the NW cone, given that any gas associated with the dust must be relatively
hot ($T_e > 10^5$K), in order to avoid the direct (local) emission line flux
dominating over the scattered emission line flux. 


\section{Conclusions}

Using high resolution optical and infrared spectroscopic observations we have
shown that a component of the narrow line gas close to the nucleus of
Cygnus A is  undergoing rotation about an axis that is close to
perpendicular to the radio axis. We also find evidence for turbulent 
broadening of the emission lines close to the nucleus, as well as 
non-circular motions and outflows in the NW cone.

The amplitude and gradient of the rotation curve close to the nucleus imply a mass of
$2.5\pm0.7\times10^9$M$_{\odot}$, and a mass density
of $4.7\pm1.3\times10^3$ M$_{\odot}$ pc$^{-3}$, within 50pc of the nucleus. 
This mass density
is orders of magnitude larger than the stellar mass densities
measured on the same scale in the  giant elliptical galaxies
of similar absolute luminosity to Cygnus A (e.g. Faber et al. 1997).

If we associate the large central mass with a supermassive
black hole, we conclude that there is nothing unusual about the mass
of the black hole in Cygnus A; the mass is entirely consistent with the
absolute brightness of the host galaxy, given the global correlations
between black hole mass and host galaxy luminosity derived for 
non-active galaxies (McClure \& Dunlop 2002). Therefore, extreme radio loudness 
and prodigious quasar-like nuclear activity are unlikely to be
solely a consequence of the large absolute masses of the central
black holes. It is probable that other factors, such as the
mass accretion rate, are at least as important as black hole
mass in determining the nature and power of the nuclear activity in
galaxies.   

Whereas the large rotational amplitude close to the nucleus supports
the black hole paradigm for powerful active nuclei, the radial outflows
detected in the NW cone bear witness to the type of AGN-induced feedback 
mechanisms which may have limited the growth of both the nuclear
black holes and the galaxy bulges in the early universe. Indeed, there
have been suggestions that such feedback leads to
the correlations between black hole mass and global galaxy properties
deduced for nearby inactive galaxies (Silk \& Rees 1998, Fabian 1999).

To our knowledge this is the first attempt to make a direct,
dynamical estimate for the mass of the black hole
in a  source harbouring a luminous, quasar-like AGN. In terms of the 
feasibility of making future direct mass estimates of the
black holes in other powerful AGN, it is clear from this study that
it is important to undertake complete velocity mapping of the near-nuclear
regions at high spatial resolution in order to distinguish
between gravitational and activity-induced gas motions. 

\subsection*{Acknowledgments} Based on observations with the NASA/ESA {\it 
Hubble Space Telescope}, obtained at the Space Telescope Science Institute,
which is operated by the Association of Universities for Research in
Astronomy (AURA), Inc., under NASA contract NAS5-26555. The W.M.
Keck Observatory is operated as a scientific partnership
between the California Institute of Technology, the
University of California, and the National Aeronautics and 
Space Administration. KAW and TR acknowledge
financial support from PPARC. We thank the anonymous referee for
useful suggestions.
{}

\end{document}